%% file: gipsy-hoil-arXiv.tex
\documentclass{easychair}

\pdfoutput=1

\input{commands}

\newcommand{\hoil}{HOIL\index{HOIL}}
\newcommand{\translucid}{TransLucid\index{TransLucid}}
\newcommand{\puredata}{PureData\index{PureData}}

\begin{document}

\title{Using the General Intensional Programming System (GIPSY) for Evaluation of Higher-Order Intensional Logic (HOIL) Expressions}

\titlerunning{Using {\gipsy} for Evaluation of {\hoil} Expressions}

\author{
	Serguei A. Mokhov\\
	{Concordia University, Montreal, Canada}\\
	{\url{mokhov@cse.concordia.ca}}\\
\and
	Joey Paquet\\
	{Concordia University, Montreal, Canada}\\
	{\url{paquet@cse.concordia.ca}}
}

\authorrunning{Mokhov and Paquet}

\maketitle

%
% Abstract
%

\begin{abstract}
The General Intensional Programming System ({\gipsy}) has been built
around the Lucid family of intensional programming languages that rely on the higher-order
intensional logic (HOIL) to provide context-oriented multidimensional reasoning
of intensional expressions. HOIL combines functional programming with various
intensional logics to allow explicit context expressions to be evaluated
as first-class values that can be passed as parameters to functions and
return as results with an appropriate set of operators defined on contexts.
{\gipsy}'s frameworks are implemented in {\java} as a collection of replaceable components
for the compilers of various Lucid dialects and the demand-driven eductive
evaluation engine that can run distributively. {\gipsy} provides support
for hybrid programming models that couple intensional and imperative languages
for a variety of needs. Explicit context expressions limit the scope of evaluation
of math expressions (effectively a Lucid program is a mathematics or physics
expression constrained by the context)
in tensor physics, regular math in multiple dimensions, etc., and for cyberforensic
reasoning as one of the use-cases of interest.
Thus, {\gipsy} is
a support testbed
for {\hoil}-based languages some of which enable such reasoning, as in formal
cyberforensic case analysis with event reconstruction. In this paper
we discuss the {\gipsy} architecture, its evaluation engine and example
use-cases.
\end{abstract}

\section{Introduction}

The {\gipsy} project is an ongoing effort aiming at providing a flexible platform
for the investigation on the intensional programming model as realized by
the latest versions of the {\lucid} programming
language~\cite{lucid76,lucid77,lucid85,lucid95,nonprocedural-iterative-lucid-77},
a multidimensional context-aware
language whose semantics is based on possible worlds semantics~\cite{kripke59,kripke69}.
{\gipsy} provides an integrated framework for compiling programs written in theoretically all
variants of {\lucid}, and even any language of intensional nature that can be translated
into some kind of ``generic Lucid'' (e.g.
{\gipl}~\cite{paquetThesis,gipsy-simple-context-calculus-08}
or {\translucid}~\cite{eager-translucid-secasa08,multithreaded-translucid-secasa08}).

Historically, the concept of {\gipsy} was conceived as a very modular collection of
frameworks geared towards sustainable support for the intensional programming
languages and embracing continuous iterative revision and development
overcoming defects of an earlier
{\glu} system~\cite{glu1,glu2,agi95glu}
that did not survive for very long
due to its inflexibility to extend to the newer dialects,
and its unmaintainability defects~\cite{paquetThesis,gipsy-arch-2000,gipsy2005}.
The {\gipsy}'s design centered around the compiler framework ({\gipc}), the eduction execution engine ({\gee}),
a.k.a the run-time execution environment, a sort of virtual machine for
execution of the intensional logic expressions, and the programming environment ({\ripe}).
The former of the three is responsible to support multiple compilers in a similar
compiler framework that all produce a consistent, well agreed on binary format,
essentially a compiled GIPSY program, as a binary output. The second performs
lazy demand-driven, potentially parallel/distributed evaluation of the compiled
Lucid programs, or as we call them, {\gipsy} programs.

\subsection{Eductive Model of Computation}

The first operational model for computing Lucid programs was designed independently
by Cargill at the University of Waterloo and May at the University of Warwick,
based directly on the formal semantics of {\lucid}, itself based on Kripke models and
possible-worlds semantics~\cite{kripke59,kripke69}. This technique was later extended
by Ostrum for the implementation of the Luthid interpreter~\cite{luthid}. Luthid being
tangential to standard {\lucid}, its implementation model was later used as a basis
to the design of the pLucid interpreter by Faustini and Wadge~\cite{eductive-interpreter}.
This program evaluation model is now called {\it eduction} and opens doors for
distributed execution of such
programs~\cite{swobodaphd04,intensionalisation-tools,distributed-context-computing,marf-gipsy-distributed-ispdc08}.

The concept of eduction can be described as ``tagged-token demand-driven dataflow''
computing (whereupon {\lucid} influenced a popular media platform and language
called {\puredata}~\cite{puredata}). The central concept to this model of execution
is the notion of generation, propagation, and consumption of {\it demands} and
their resulting {\it values}.
Lucid programs are declarative programs where every identifier is defined as
a {\hoil} expression using other identifiers and an underlying algebra.
An initial demand for the value of a certain identifier is generated,
and the eduction engine, using the defining expression of this identifier,
generates demands for the constituting identifiers of this expression,
on which operators are applied in their embedding expressions.
These demands in turn generate other demands, until some demands eventually
evaluate to some values, which are then propagated back in the chain of demands,
operators are applied to compute expression values, until eventually the value
of the initial demand is computed and returned.

Lucid identifiers and expressions inherently vary in a {\it multidimensional context space},
i.e. any identifier or expression can be evaluated in a multidimensional context,
thus leading to have identifiers and expressions representing a set of values,
one value for each possible context in which the identifier or expression can be evaluated.
This is brining the notion of {\em intensionality}, where identifiers are defined by
intensional expressions i.e. expressions whose evaluation varies in a multidimensional
context space, which can then be constrained by a particular multidimensional
context specification. Note that Lucid variables and expressions represent
``dimensionally abstract'' concepts, i.e. they do not explicitly mention their dimensionality.
For example, Newton's Law of Universal Gravitation can be written literally in {\lucid} as:

%\vspace{3mm}
\noindent{\tt F = (G * m1 * m2) / r * r;}
%\vspace{3mm}

\noindent and can then be evaluated in different dimensional manifolds
(i.e. $n$-dimensional spaces), keeping the same definition, but being evaluated
in contexts varying in their dimensionality.
For example, {\tt F} can be evaluated in a one-dimensional space,
yielding a single scalar, or in a three-dimensional manifold, yielding a
three-dimensional vector.
Note that a {\tt time} dimension could also be added where, for example,
the masses ({\tt m1} and {\tt m2}) and/or the distance between them ({\tt r)}
might be defined as to vary in time. In such a case, the expression would then
inherently be varying in the time dimension, due to some of its constituents
varying in this dimension.

\subsection{Intensional Logic and Programming}
\index{intensional!logic}
\index{intensional!programming}

Intensional programming, in the sense of the latest evolutions of {\lucid},
is a programming language paradigm based on the notion of declarative programming
where the declarations are evaluated in an inherent multidimensional context space.
The context space being in most cases infinite,
intensional programs are evaluated using a lazy demand-driven model of execution
that we introduced earlier -- {\it eduction}~\cite{eductive-interpreter},
where the program identifiers are evaluated in a restricted context space,
in fact, a {\it point} in space, where each demand is generated, propagated,
computed and stored as an {\it identifier-context} pair~\cite{bolu04}.

Many problem domains are intensional in nature,
e.g. computation of differential and tensor equations~\cite{paquetThesis},
temporal computation and temporal
databases~\cite{paquet-intensional-databases-95,active-functional-idatabase},
multidimensional signal processing~\cite{agi95glu},
context-driven computing~\cite{crr05},
constraint programming~\cite{wanphd06},
negotiation protocols~\cite{wanphd06},
automated reasoning in cyberforensics~\cite{flucid-imf08,flucid-isabelle-techrep-tphols08,marf-into-flucid-cisse08},
multimedia and pattern recognition~\cite{marfl-context-secasa08}
among others.
The current mainstream programming languages are not well adapted
for the natural expression of the intensional aspects of such problems,
requiring the expression of the intensional nature of the problem statement
into a procedural (and therefore sequential) approach in order to provide a computational solution.

Intensional programming can be used to solve widely diversified problems,
which can be expressed using diversified languages of intensional nature.
There also has been a wide array of flavors of Lucid languages developed
over the years. Yet, very few of these languages have made it to the
implementation level. The {\gipsy} project aims at the creation of
a programming environment encompassing compiler generation for all flavors
of {\lucid}, a generic run-time system enabling the execution of programs
written in all flavors of {\lucid}. Our goal is to provide a flexible
platform for the investigation on programming languages of intensional nature,
in order to prove the applicability of intensional programming to solve
important problems.

Intensional programming is based on intensional (or multidimensional)
logics\index{intensional!logic}\index{logic!intensional}, which, in turn,
are based on natural language understanding (aspects, such as, time, belief,
situation, and direction are considered).
Intensional programming brings in {\em dimensions}\index{dimensions}
and {\em context}\index{context} to programs (e.g. space and time
in physics or chemistry). Intensional logic adds dimensions to logical
expressions; thus, a non-intensional logic\index{logic!non-intensional} can
be seen as a constant or a snapshot in all possible dimensions.
{\em Intensions are dimensions} at which a
certain statement is true or false (or has some other than a Boolean value).
{\em Intensional operators}\index{intensional!operators} are operators that
allow us to navigate within these dimensions~\cite{paquetThesis}.

\subsubsection{Temporal Intensional Logic Example}
\index{logic!temporal}

Temporal intensional logic is an extension of temporal logic that allows
to specify the time in the future or in the past~\cite{paquetThesis}.

(1) \tab{20} $E_1$ := it is raining {\bf here} {\bf today}

Context: \{\texttt{place:}{\bf here}, \texttt{time:}{\bf today}\}

(2) \tab{20} $E_2$ := it was raining {\bf here} {\it before}({\bf today}) = {\it yesterday}

(3) \tab{20} $E_3$ := it is going to rain {\it at} (altitude {\bf here} + 500 m) {\it after}({\bf today}) = {\it tomorrow}

Let's take $E_1$ from (1) above. The context is a collection of the dimensions
\api{place} and \api{time} with the corresponding tag values of {\bf here} and {\bf today}.
Then
let us fix {\bf here} to {\bf Montreal} and assume it is a {\it constant}.
In the month of May, 2009, with granularity of day, for every day, we can
evaluate $E_1$ to either {\it true} or {\it false},
as shown in \xf{fig:tag-values-1d-example}.
\begin{figure}[htb!]
\hrule
\small
\centering
\begin{verbatim}

Tags days in May:      1 2 3 4 5 6 7 8 9 ...
Values (raining?):     F F T T T F F F T ...
\end{verbatim}
\normalsize
\hrule
\caption{1D example of tag-value contextual pairs.}
\label{fig:tag-values-1d-example}
\end{figure}
If one starts varying the {\bf here} dimension (which could even be broken down
to $X$, $Y$, $Z$), one gets a two-dimensional (or 4D respectively) evaluation of $E_1$,
as shown in \xf{fig:tag-values-2d-example}.
\begin{figure}[htb!]
\hrule
\small
\centering
\begin{verbatim}

Place/Time  1 2 3 4 5 6 7 8 9 ...
Montreal    T F T F T F T F T ...
Honolulu    F F T T T F F F T ...
New York    F F F F T T T F F ...
Tampa       F T T T T T F F F ...
\end{verbatim}
\normalsize
\hrule
\caption{2D example of tag-value contextual pairs.}
\label{fig:tag-values-2d-example}
\end{figure}
Even with these toy examples we can immediately illustrate the hierarchical notion
of the dimensions in the context: so far the place and time we treated as
atomic values fixed at days and cities. In some cases, we need finer
subdivisions of the context evaluation, where time can become fixed at
hour, minute, second and finer values, and so is the place broken down
into boroughs, regions, streets, etc. and finally the $X,Y,Z$ coordinates
in the Euclidean space with the values of millimeters or finer. This notion
becomes more apparent and important e.g. in {\flucid}, a forensic
case specification language for automated reasoning in cybercrime and other
investigations.

\subsection{{\hoil}}

To summarize, expressions written in virtually all {\lucid} dialects are
correspond to higher-order intensional logic ({\hoil})
expressions with some dialect-specific instantiations. They all can
alter the context of their evaluation given a set of operators
and in some cases types of contexts, their range, and so on.
{\hoil} combines functional programming and intensional logics,
e.g. temporal intensional logic mentioned earlier. The contextual
expression can be passed as parameters and returned as results
of a function and constitute the multi-dimensional constraint
on the Lucid expression being evaluated. The corresponding
context calculus~\cite{wanphd06,gipsy-simple-context-calculus-08,tongxinmcthesis08}
defines a comprehensive set of context operators, most of which 
are set operators and the baseline operators are \api{@}
and \api{\#} that allow to switch the current context or query 
it, respectively. Other operators allow defined a context space
and a point in that context corresponding to the current context.
The context can be arbitrary large in its rank.
The identified variables of the dimension type within the context
can take on any data type, e.g. an integer, or a string, during
lazy binding of the resulting context to a dimension identifier.

\section{{\gipsy}'s Architecture}

{\gipsy} evolved from a modular collection of frameworks for
local execution into a multi-tier architecture~\cite{gipsy-multi-tier-secasa09}.
With the bright but short-lived story of {\glu} in mind, efforts were
made to design a new system with similar capacities,
but with more flexibility in mind. The new system would have to be able
to cope with the fast evolution and diversity of the Lucid family of languages,
thus necessitating a flexible compiler architecture, and a language-independent
run-time system for the execution of Lucid programs.
The architecture of the GIPSY compiler, the General Intensional Programming Compiler (GIPC)
is framework-based, allowing the modular development of compiler components
(e.g. parser, semantic analyzer and translator). It is based on the notion of
the Generic Intensional Programming Language ({\gipl}), a core language into which
all other flavors of the Lucid language can be translated to.
The notion of a generic language also solved the problem of language-independence
of the run-time system by allowing a common representation for all
compiled programs, the Generic Eduction Engine Resources ({\geer}),
which is a dictionary of run-time resources compiled from a GIPL program,
that had been previously generated from the original program using semantic
translation rules defining how the original Lucid program can be
translated into the {\gipl}. For a more complete description of the GIPSY compiler
framework, see~\cite{aihuawu02,chunleiren02,wuf04,mokhovmcthesis05}.
The architecture necessitates the presence of the
intensional-imperative type system and support
links to imperative languages being presented elsewhere~\cite{gipsy-type-system-c3s2e09}.

\subsection{General Intensional Program Compiler ({\gipc})}

The {\gipsy}, conceptually represented  at the high level in \xf{fig:gipsy}.
The type abstractions and implementations are located
in the \api{gipsy.lang} package and serve as a glue between the
compiler (known as the {\gipc} -- a General Intensional Program Compiler) and the
run-time system (known as the {\gee} -- a General Eduction Engine) to do the static
and dynamic semantic analyses and evaluation respectively. The {\gipsy}
is a very modular system allowing most components to be replaceable
as long as they comply with some general architectural interface or API.
One of such API interfaces is the \api{GIPSYProgram} (conceptually
represented as {\geer} -- GEE Resource -- a dictionary of run-time resources) that contains
among other things the type annotations that can be statically inferred
during compilation. At run-time, the engine does it's own type checking
and evaluation when traversing the AST stored in the GEER and evaluating expressions represented
in the tree. Since both the {\gipc} and the {\gee} use the same type system
to do their analysis, they consistently apply the semantics and rules
of the type system with the only difference that the {\gee}, in addition to the
type checks, does the actual evaluation.
The {\gipsy} is primarily implemented in {\java}.

%\begin{figure}
%	\begin{centering}
%	\includegraphics[width=.6\textwidth]{Dgeneral}
%	\caption{General Intensional Programming System Overview.}
%	\label{fig:gipsy}
%	\end{centering}
%\end{figure}

\begin{figure}[htb!]
	\begin{centering}
	\includegraphics[width=.5\textwidth]{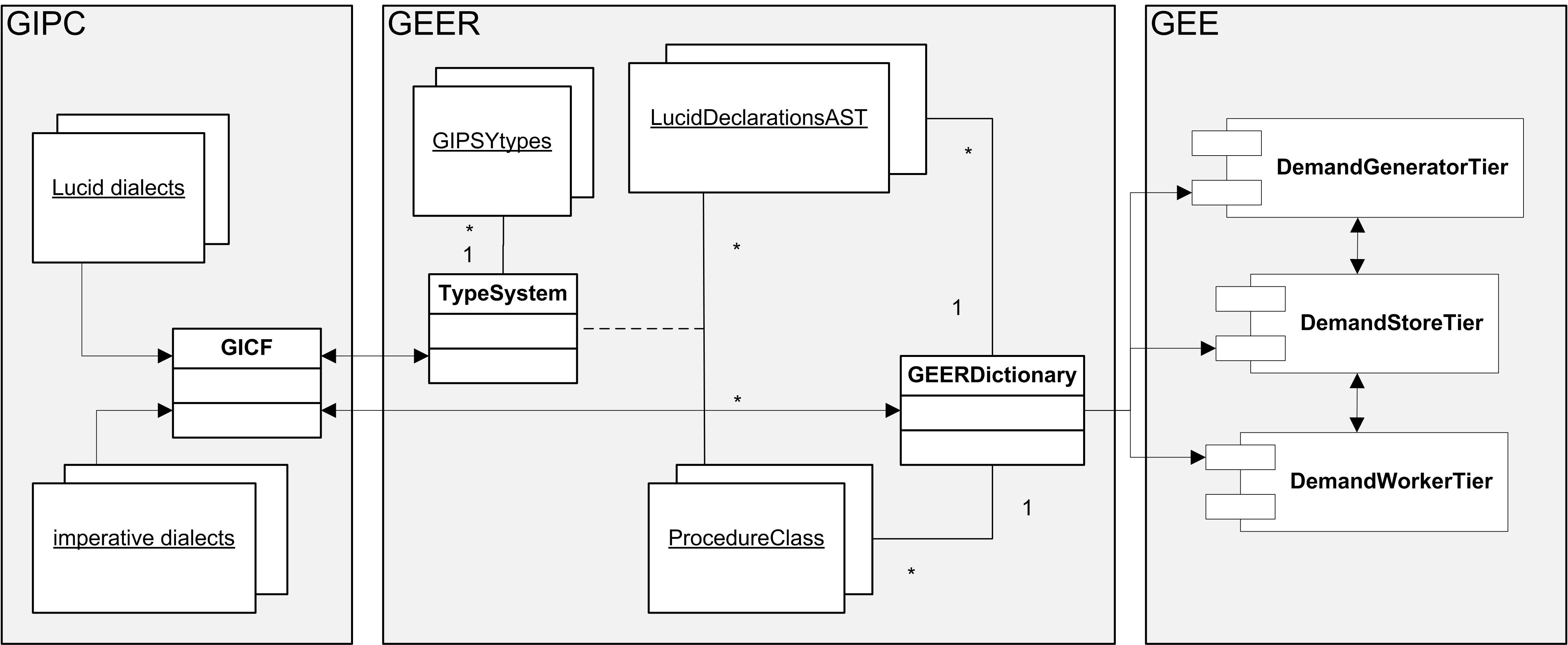}
	\caption{{\gipsy}'s {\gipc}-to-{\gee} {\geer} Flow Overview in Relation to the GIPSY Type System.}
	\label{fig:gipsy}
	\end{centering}
\end{figure}

% \subsubsection{GIPC Preprocessor}
% \label{sect:gipc-preprocessor}
% \index{Preprocessor}
% \index{GIPC!Preprocessor}
% \index{Preprocessor!GIPC}

%\begin{figure}
%	\begin{centering}
%	\includegraphics[width=\textwidth]{gipc-preprocessor}
%	\caption{GIPC Framework with the Preprocessor}
%	\label{fig:gipc-preprocessor}
%	\end{centering}
%\end{figure}

\begin{figure}[htb!]
	\begin{centering}
	\includegraphics[width=.5\textwidth]{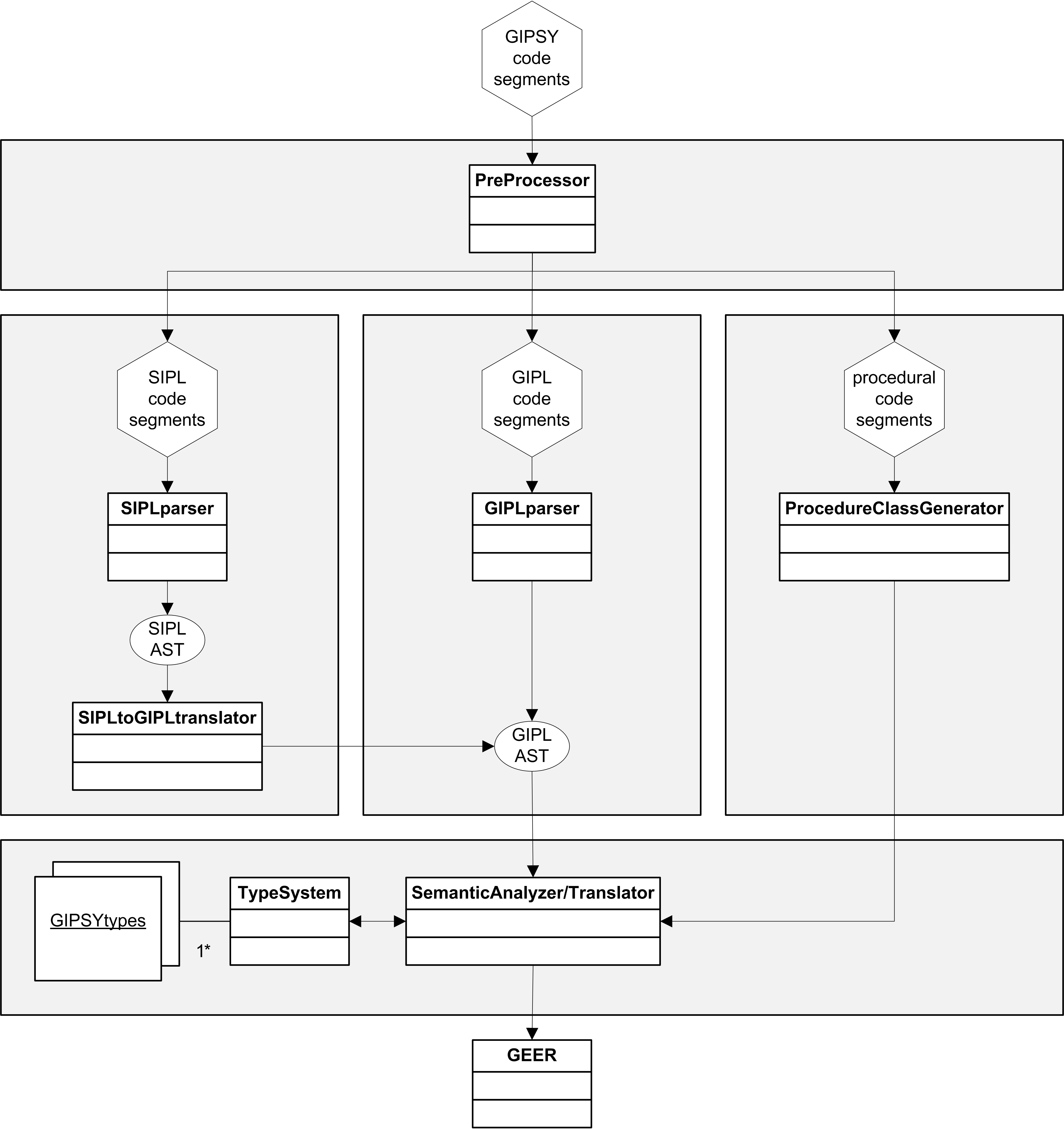}
	\caption{GIPC Framework}
	\label{fig:gipc-preprocessor}
	\end{centering}
\end{figure}

The \api{Preprocessor}~\cite{mokhovmcthesis05,mokhovgicf2005}
is something that is invoked first by the GIPC (see \xf{fig:gipc-preprocessor})
on incoming GIPSY program's source code stream. The \api{Preprocessor}'s
role is to do preliminary program analysis, processing, and splitting
the source GIPSY program into ``chunks'', each written in a different
language and identified by a {\it language tag}. In a very general view,
a GIPSY program is a hybrid program consisting of different languages
in one or more source file; then, there has to be an interface between
all these code segments. Thus, the \api{Preprocessor} after some initial
parsing (using its own preprocessor syntax) and producing the initial parse tree,
constructs a preliminary dictionary of symbols used throughout the program.
This is the basis for type matching and semantic analysis applied later on.
This is also where the first step of type assignment occurs, especially
on the boundary between typed and typeless parts of the program,
e.g. {\java} and a specific Lucid dialect.
The \api{Preprocessor} then splits the code segments of the GIPSY
program into chunks preparing them to be fed to the respective
concrete compilers for those chunks. The chunks are represented through
the \api{CodeSegment} class that the \api{GIPC} collects.

\subsection{General Eduction Engine ({\gee})}

The design architecture adopted is a distributed multi-tier architecture, where each tier can have any number of instances. The architecture bears resemblance with a peer-to-peer architecture, e.g.:

\begin{itemize}
\item Demands are propagated without knowing where they will be processed or stored.
\item Any tier or node can fail without the system to be fatally affected.
\item Nodes and tiers can seamlessly be added or removed on the fly as computation is happening.
\item Nodes and tiers can be affected at run-time to the execution of any GIPSY program, i.e. a specific node or tier could be computing demands for different programs.
\end{itemize}

\subsubsection{Generic Eduction Engine Resources}

One of the central concepts of our solution is {\it language independence} of the run-time system. In order to achieve that, we rely on an intermediate representation that is generated by the compiler: the Generic Eduction Engine Resources (GEER). The General Intensional Programming Compiler (GIPC) compiles a program into an instance of  the GEER, a dictionary of identifiers compiled from the program by the compiler~\cite{wuf04,mokhovmcthesis05}. The compiler framework provides with the potential to allow the easy addition of any flavor of the Lucid language to be added through automated compiler generation taking semantic translation rules in input~\cite{aihuawu02}.
As the name suggests, the GEER structure is generic, in the sense that the data structure and semantics of the GEER are independent of the language in which its corresponding source code was written. This is necessitated by the fact that the engine was designed to be ``source language independent'', an important feature made possible by the presence of the Generic Intensional Programming Language (GIPL) as a generic language in the Lucid family of languages.
Thus, the compiler first translates the source program (written in any flavor of Lucid) into ``generic Lucid'', then generate the GEER run-time resources for this program, which is then made available at run-time to the various tiers upon demand.
The GEER contains, for all Lucid identifiers in a given program, typing information, rank i.e. dimensionality information, as well as an abstract syntax tree representation of the declarative definition of the identifier.
It is this latter tree that is later on traversed by the demand generator in order to proceed with demand generation. In the case of hybrid Lucid programs, the GEER also contains a dictionary of procedures called by the Lucid program, known as Procedure Classes, as they in fact are {\it wrapper classes} wrapping procedures inside a Java class in cases where the functions being called are not written in Java~\cite{mokhovmcthesis05,mokhovgicf2005}.

\subsubsection{GIPSY Tier}

The architecture adopted for this new evolution of the GIPSY is a multi-tier architecture where the execution of GIPSY programs is divided in three different tasks assigned to separate tiers.
%, as described in the following paragraphs.
Each GIPSY tier is a separate process that communicates with other tiers using {\it demands}, i.e. the GIPSY Multi-Tier Architecture operational mode is fully {\it demand-driven}. The demands are generated by the tiers and migrated to other tiers using the Demand Store Tier. In this paper, we refer to a ``tier'' as an abstract and generic entity that represents a computational unit independent of other tiers and that collaborates with other tiers to achieve program execution as a group.

\subsubsection{GIPSY Node}

Abstractly, a GIPSY node is a computer that has registered for the hosting of one or more GIPSY tier.
GIPSY Nodes are registered through a GIPSY Manager instance. Technically, a GIPSY Node is a controller
that wraps GIPSY Tier instances, and that is remotely reporting and being controlled
by a GIPSY Manager. %(see below).
Operationally, a GIPSY Node hosts one tier Controller for each kind of tier~(see \xf{fig:GIPSYnodeDesign}). The Tier Controller acts as a factory that will, upon necessity, create {\it instances} of this tier, which provide the concrete operational features of the tier in question. This model permits scalability of computation by allowing the creation of new tiers instances as existing tier instances get overloaded or lost.

\begin{figure}[htb!]
	\begin{centering}
	\includegraphics[width=.5\textwidth]{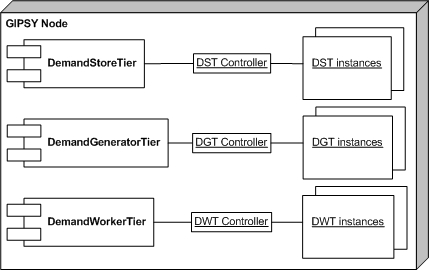}
	\caption{Design of the GIPSY Node}
	\label{fig:GIPSYnodeDesign}
	\end{centering}
\end{figure}

\subsubsection{GIPSY Instance}

A GIPSY Instance is a set of interconnected GIPSY Tiers deployed on GIPSY Nodes executing GIPSY programs by sharing their respective GEER instances. A GIPSY Instance can be executing across different GIPSY Nodes, and the same GIPSY Node may host GIPSY Tiers that a part of separate GIPSY Instances

\subsubsection{Demand Generator Tier}

The Demand Generator Tier (DGT) generates demands according to the program declarations and definitions stored in one of the instances of GEER that it hosts. The demands generated by the Demand Generator Tier instance can be further processed by other Demand Generator Tiers instances (in the case of intensional demands) or Demand Worker Tier instances (in the case of procedural demands), the demands being migrated across tier instances through a Demand Store Tier instance. Each DGT instance hosts a set of GEER instances that corresponds to the Lucid programs it can process demands for. A demand-driven mechanism allows the Demand Generator Tier to issue system demands requesting for additional GEER instances to be added to its GEER Pool, thus enabling DST instances to process demands for additional programs as they are executed on the GIPSY instances they belong to.

\subsubsection{Demand Store Tier}

The Demand Store Tier (DST) acts as a tier middleware in order to migrate demands between tiers. In addition to the migration of the demands and values across different tiers, the Demand Store Tier provide persistent storage of demands and their resulting values, thus achieving better processing performances by not having to re-compute the value of every demand every time it is re-generated after having been processed. From this latter perspective, it is equivalent to the historical notion of {\it warehouse} in the eduction model of computation. A centralized communication point or warehouse is likely to become an execution bottleneck. In order to avoid that, the Demand Store Tier uses a peer-to-peer architecture and mechanism to connect all Demand Store Tier instances in a given GIPSY instance. This allows any demand or its resulting value to be stored on any DST instance, but yet allows abstract querying for a specific demand value on any of the DST instances. If the demanded value is not found on the DST instance receiving the demand, it will contact its DST instance peers using a peer-to-peer mechanism. This mechanism allows to see the Demand Store abstractly as a single store that is, behind the scenes, a distributed one.

\subsubsection{Demand Worker Tier}

The Demand Worker Tier (DWT) processes {\it procedural demands} i.e. demands for the execution of functions or methods defined in a procedural language, which are only present in the case where hybrid intensional programs are being executed. The DGT and DWT duo is an evolution of the generator-worker architecture adopted in GLU~\cite{glu1,glu2}. It is through the operation of the DWT that increased granularity of computation is achieved. Similarly to the DGT, each DWT instance hosts a set of compiled procedures (Procedure Classes) that corresponds to the procedural demands it can process. A demand-driven mechanism allows the Demand Worker Tier to issue system demands requesting for additional Procedure Classes to be added to its Procedure Class Pool, thus achieving increasing capacities over time, on demand.

\subsubsection{GIPSY Instance Manager}

A GIPSY Instance Manager (GIM) is a component that enables the registration of GIPSY Nodes and Tiers, and to allocate them to the GIPSY Instances that it manages. The GIPSY Instance Manager interacts with the allocated tiers in order to determine if new tiers and/or nodes are necessary to be created, and issue demands to GIPSY Nodes to spawn new tier instances if need there be. In order to ease the node registration, the GIPSY Instance Manager tier can be implemented using a web interface, so that users can register nodes using a standard web browser, rather than requiring a client. GIPSY Instance Managers are peer-to-peer components, i.e. users can register a node through any GIPSY Instance Manager, which will then inform all the others of the presence of the new node, which will then be available for hosting new GIPSY Tiers at the request of any of the GIPSY Instance Managers currently running. The GIM uses {\it system demands} to communicate with Nodes and Tiers.

\section{Context-Oriented Reasoning}
\label{sect:context-reasoning}

As mentioned earlier, the reasoning aspect of {\gipsy} is a particularity
of a {\lucid} dialect rather than the architecture, and in this paper we
look at it from the reasoning angle. The architecture is general enough
to go beyond reasoning -- in the essence it is an evaluation of intensional
logic expressions. Now if those expressions form a language dialect, that
helps us with reasoning, such as {\flucid} to reason about cybercrime
incidents and claims. Some other {\lucid} dialects have less relevance
in that regard, but still can be included into the system -- such as
{\tlucid} for tensor fields evaluation of particles in plasma~\cite{paquetThesis},
reactive programming, multi-core processing~\cite{eager-translucid-secasa08,multithreaded-translucid-secasa08},
software versioning, and others.

\subsection{{\lucx}}

{\lucx}~\cite{kaiyulucx,wanphd06} is a fundamental extension of {\gipl} and the {\lucid} family as a whole that promotes the contexts as first-class values thereby creating a ``true'' generic {\lucid} language. Wan in~\cite{kaiyulucx,wanphd06} defined a new collection of set operators (e.g. union, intersection, box, etc.) on the multidimensional contexts, which will help with the multiple explanations of the evidential statements in forensic evaluation where the context sets are often defined as cross products (boxes), intersections, and unions. Its further specification, refinement, and
implementation details are presented in~\cite{tongxinmcthesis08,gipsy-simple-context-calculus-08}.

\subsection{Operational Semantics for Reasoning about Lucid Expressions}
\label{sect:semantics}
\label{appdx:semantics}

Here for convenience we provide the semantic rules of {\ilucid}~\cite{paquetThesis}
(see \xf{fig:gipl-semantics}), {\lucx}~\cite{wanphd06} (see \xf{fig:lucx-semantics}).

\begin{figure}[htb!]
\scriptsize
% \begin{eqnarray*}
\begin{eqnarray}
{\mathbf{E_{cid}}} &:& \frac
  {\johndef(\myid)=(\texttt{const},c)} {\context{\myid}{c}}\\\nonumber\\
{\mathbf{E_{opid}}} &:& \frac
  {\johndef(\myid)=(\texttt{op},f)}
  {\context{\myid}{\myid}}\\\nonumber\\
{\mathbf{E_{did}}} &:& \frac
  {\johndef(\myid)=(\texttt{dim})}
  {\context{\myid}{\myid}}\\\nonumber\\
{\mathbf{E_{fid}}} &:& \frac
  {\johndef(\myid)=(\texttt{func},\myid_i,E)}
  {\context{\myid}{\myid}}\\\nonumber\\
{\mathbf{E_{vid}}} &:& \frac
  {\johndef(\myid)=(\texttt{var},E)\qquad
   \context{E}{v}}
  {\context{\myid}{v}}\\\nonumber\\
{\mathbf{E_{c_T}}} &:& \frac
  {\context{E}{\textit{true}}\qquad
   \context{E'}{v'}
  }
  {\context{\myifthenelse}{v'}}\\\nonumber\\
{\mathbf{E_{c_F}}} &:& \frac
  {\context{E}{\textit{false}}\qquad
   \context{E''}{v''}
  }
  {\context{\myifthenelse}{v''}}\\\nonumber\\
{\mathbf{E_{tag}}} &:& \frac
  {\context{E}{\myid}\qquad
   \johndef(\myid)=(\texttt{dim})
  }
  {\context{\#E}{\mathcal{P}(\myid)}}\\\nonumber\\
{\mathbf{E_{at}}} &:& \frac
  {\context{E'}{\myid}\qquad
   \johndef(\myid)=(\texttt{dim})\qquad
   \context{E''}{v''}\qquad
   \mathcal{D},\mathcal{P}\mydagger[\myid\mapsto v''] \vdash E : v
  }
  {\context{E\;@E'\;E''}{v}}\\\nonumber\\
{\mathbf{E_{w}}} &:& \frac
  {\qcontext{Q}{\mathcal{D}',\mathcal{P}'}\qquad
   \mathcal{D}',\mathcal{P}' \vdash E : v
  }
  {\context{E\;\mathtt{where}\;Q}{v}}\\\nonumber\\
{\mathbf{Q_{dim}}} &:& \frac
  {}
  {\qcontext{\texttt{dimension}\;\myid}
   {\mathcal{D}\mydagger[\myid\mapsto(\texttt{dim})],
   \mathcal{P}\mydagger[\myid\mapsto 0]}
  }\\\nonumber\\
{\mathbf{Q_{id}}} &:& \frac
  {}
  {\qcontext{\myid=E}
   {\mathcal{D}\mydagger[\myid\mapsto(\texttt{var},E)],
    \mathcal{P}}
  }\\\nonumber\\
{\mathbf{QQ}} &:& \frac
  {\qcontext{Q}{\mathcal{D}',\mathcal{P}'}\qquad
   \mathcal{D}',\mathcal{P}' \vdash Q' : \mathcal{D}'',\mathcal{P}''
  }
  {\qcontext{Q\;Q'}{\mathcal{D}'',\mathcal{P}''}}\\\nonumber\\
{\mathbf{E_{op}}} &:& \frac
  {\context{E}{\myid}\qquad
   \johndef(\myid)=(\texttt{op},f)\qquad
   \context{E_i}{v_i}
  }
  {\context{E(E_1,\ldots,E_n)}{f(v_1,\ldots,v_n)}}\\\nonumber\\
{\mathbf{E_{fct}}} &:& \frac
  {\context{E}{\myid}\qquad
   \johndef(\myid)=(\texttt{func},\myid_i,E')\qquad
   \context{E'[\myid_i\leftarrow E_i]}{v}
  }
  {\context{E(E_1,\ldots,E_n)}{v}}
% \end{eqnarray*}
\end{eqnarray}
\normalsize
\caption{Extract of Operational Semantics Rules of {\gipl}}
\label{fig:gipl-semantics}
\end{figure}

\begin{figure}[htb!]
\scriptsize
% \begin{eqnarray*}
\begin{eqnarray}
% Lucx
% \hline\\
{\mathbf{E_{\#}}} &:& \frac
  {}
  {\context{\#}{\mathcal{P}}}\\\nonumber\\
{\mathbf{E_{.}}} &:& \frac
  {\context{E_{2}}{\myid_{2}}\qquad
   \johndef(\myid_{2})=(\texttt{dim})
  }
  {\context{E_{1}.E_{2}}{\mathit{tag }(E_{1} \downarrow \{\myid_{2}\})}
  }\\\nonumber\\
{\mathbf{E_{tuple}}} &:& \frac
  {\context{E}{\myid}\qquad
   \mathcal{D}\mydagger[\myid \mapsto (\texttt{dim})]\qquad
   \mathcal{P}\mydagger[\myid \mapsto 0]\qquad
   \context{E_{i}}{v_{i}}
  }
  {\context{\langle E_{1},E_{2},\ldots,E_{n}\rangle E}{v_{1} \;\; \mathit{fby}.\myid \;\; v_{2} \;\; \mathit{ fby}.\myid \;\; \ldots \;\; v_{n} \;\; \mathit{fby}.\myid \;\; \texttt{eod}}
  }\\\nonumber\\
{\mathbf{E_{select}}} &:& \frac
  {E=[\texttt{d}:\texttt{v'}]\qquad
   E' = \langle \texttt{E}_{1},\ldots,\texttt{E}_{n} \rangle\texttt{d}
   \mathcal{P'}=\mathcal{P}\mydagger[d \mapsto v']\qquad
   \mathcal{D},\mathcal{P}' \vdash E' : v
  }
  {\context{\mathit{select}(E,E')}{v}
  }\\\nonumber\\
{\mathbf{E_{at(c)}}} &:& \frac
  {\context{\mathcal{C}}{\mathcal{P}}\qquad
   \context{E}{v}
  }
  {\context{E\;@C}{v}}\\\nonumber\\
{\mathbf{E_{at(s)}}} &:& \frac
  {\context{\mathcal{C}}{\{\mathcal{P}_{1},\ldots,\mathcal{P}_{2}\}}\qquad
   \mathcal{D},\mathcal{P}_{i:1 \ldots m} \vdash E : v_{i}
  }
  {\context{E\;@C}{\{v_{1},\ldots,v_{m}\}}}\\\nonumber\\
{\mathbf{C_{context}}} &\!\!\!\!\!\!\!\!\!\!:\!\!\!\!\!\!\!\!\!\!& \frac
   {
    \begin{array}{l}
    \context{E_{d_{j}}}{\myid_{j}}\qquad
    \johndef(\myid_{j})=(\texttt{dim})\\
    \context{E_{i_{j}}}{v_{j}}\qquad
    \mathcal{P}' = \mathcal{P}_{0}\mydagger[\myid_{1}\mapsto v_{1}]\mydagger \ldots \mydagger [\myid_{n}\mapsto v_{n}]
    \end{array}
   }
   {
    \context{[E_{d_{1}}:E_{i_{1}},E_{d_{2}}:E_{i_{2}},\ldots,E_{d_{n}}:E_{i_{n}}]}{\mathcal{P}'}
   }\\\nonumber\\
{\mathbf{C_{box}}} &\!\!\!\!\!\!\!\!\!\!:\!\!\!\!\!\!\!\!\!\!& \frac
   {
    \begin{array}{l}
    \context{E_{d_{i}}}{\myid_{i}}\qquad\qquad\qquad
    \johndef(\myid_{i})=(\texttt{dim})\\
    \{E_1,\ldots,E_n\} = \mathit{dim}(\mathcal{P}_1)=\ldots=\mathit{dim}(\mathcal{P}_m)\\
    E' = \texttt{f}_{p}(\texttt{tag}(\mathcal{P}_1),\ldots,\texttt{tag}(\mathcal{P}_m))\qquad
    \context{E'}{\mathit{true}}
    \end{array}
   }
   {
    \context{\mathit{Box}[E_1,\ldots,E_n|E']}{\{\mathcal{P}_1,\ldots,\mathcal{P}_m\}}
   }\\\nonumber\\
{\mathbf{C_{set}}} &:& \frac
  {\context{E_{w:1 \ldots m}}{\mathcal{P}_m}
  }
  {\context{\{E_1,\ldots,E_m\}}{\{\mathcal{P}_1,\ldots,\mathcal{P}_w\}}
  }\\\nonumber\\
{\mathbf{C_{op}}} &:& \frac
  {
   \context{E}{\myid}\qquad
   \johndef(\myid)=(\texttt{cop}, f)\qquad
   \context{C_i}{v_i}
  }
  {\context{E(C_1,\ldots,C_n)}{f(v_1,\ldots,v_n)}
  }\\\nonumber\\
{\mathbf{C_{sop}}} &:& \frac
  {
   \context{E}{\myid}\qquad
   \johndef(\myid)=(\texttt{sop}, f)\qquad
   \context{C_i}{\{v_{i_{1}},\ldots,v_{i_{k}}\}}
  }
  {\context{E(C_1,\ldots,C_n)}{f(\{v_{1_{1}},\ldots,v_{1_{s}}\},\ldots,\{v_{n_{1}},\ldots,v_{n_{m}}\})}
  }%\\\nonumber\\
% \end{eqnarray*}
\end{eqnarray}
\normalsize
\caption{Extract of Operational Semantics of {\lucx}}
\label{fig:lucx-semantics}
\end{figure}

\subsection{Higher Order Context}

HOCs represent essentially nested contexts, e.g.
as conceptually shown in \xf{fig:nested-context-concept}
modeling evidential statement for forensic specification
evaluation. Such a context representation can be modeled
as a tree in an OO ontology or a context-set as in {\lucx}.
The early notion and specification of nested context first
appeared
Swoboda's works~\cite{swobodaphd04,intensionalisation-tools,distributed-context-computing},
but there the evaluation has taken place only at the leaf context
nodes. Another, more recent work on the configuration specification
as a context in the intensional manner was the
MARFL language~\cite{marfl-context-secasa08,marf-into-flucid-cisse08},
allowing evaluation at arbitrary nesting of the configuration
context with some defaults in the intermediate and leaf context nodes.

\begin{figure*}[ht!]
	\includegraphics[width=\textwidth]{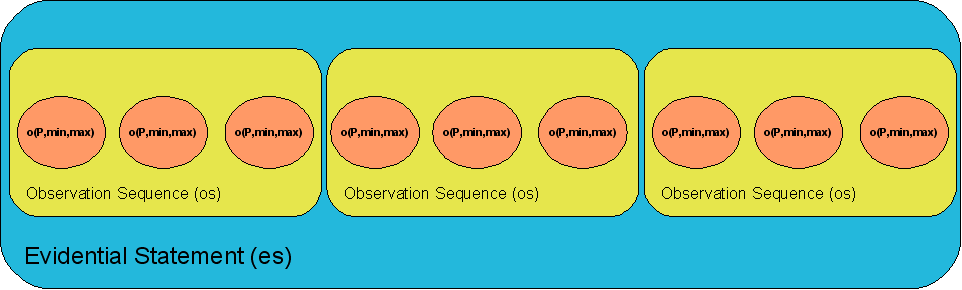}
	\caption{Nested Context Hierarchy Example for Cyberforensic Investigation}
	\label{fig:nested-context-concept}
\end{figure*}

\subsection{Reasoning About Cyberforensic Cases and {\flucid}}

A {\lucid} dialect, {\flucid}~\cite{flucid-imf08,flucid-isabelle-techrep-tphols08} develops a specification of
a cyber incident for analysis of claims of witnesses against
encoded evidential statements to see if they agree or not
and if they do provide potential backtraces of event
reconstruction.
The hierarchical context space similar
to the one defined in the previous section is also used in {\flucid}
to denote a context of a cybercrime case as a hierarchy consisting
of the evidential statement $es$, that consists of observation
sequences (``stories'' told by evidence and witnesses) $os$, which in
turn consist of observations $o$, that denote a certain observer
property $P$ and its duration $[\min, \min+\max]$, as shown in 
\xf{fig:nested-context-concept}.
{\flucid} draws from an earlier formal
approach using finite state automata by Gladyshev~\cite{blackmail-case,printer-case}
that is not very usable and all of the benefits of Lucid
and intensional evaluation with functions and operators
that navigate withing the higher-order context space of
evidence and witness stories to evaluate claims. The
Demptser-Shafer theory~\cite{prob-argumentation-systems,shafer-evidence-theory}
is used to assigned weights, such
as credibility and admissibility to witnesses evidence
as a part of reasoning parameters.

\section{Conclusion}

We presented a modular intensional programming research platform, {\gipsy}, for reasoning tasks
of HOIL expressions.
The concept of context as a first-class
value is central in the programming paradigms {\gipsy} is build to explore,
such as a family of the {\lucid} programming languages. At the time
of this writing {\gipsy} has support for compilation of {\gipl},
{\ilucid}, {\lucx}, {\jlucid}, and {\olucid} and the execution of if the
former two with the other being completed. The DMS for distributed
transport of the demands has implementations in Jini, plain RMI,
and JMS.

\subsection{Future Work}

The future and ongoing work within the context of {\gipsy} is a complete
formalization of its hybrid intensional-imperative
type system~\cite{gipsy-type-system-c3s2e09}, the revision of the syntax
and semantics of the {\flucid} language, and the multi-tier
overhaul of the evaluation engine ({\gee}) including support
for OO intensional dialects.

\subsection{Acknowledgments}

This work was funded by NSERC and the Faculty of Computer Science and Engineering,
Concordia University. Thanks to many of the GIPSY project team members for their
valuable contributions, suggestions, and reviews, including Peter Grogono,
Emil Vassev, Xin Tong, and Amir Pourteymour.

%%%%%%%%%%%%%%%%%%%%%%%%%%%%%%%%%%%%%%
% Refs:
%
\label{sect:bib}
\bibliographystyle{abbrv}
\bibliography{gipsy-hoil-arXiv}

\end{document}

%% file: commands.tex
\newcommand{\xf}[1]{Figure~\ref{#1}}

\newcommand{\gipc}{{GIPC\index{GIPC}}}

\newcommand{\gee}{{GEE\index{GEE}}}
\newcommand{\geer}{{GEER\index{GEER}}}
\newcommand{\gipsy}{{GIPSY\index{GIPSY}}}
\newcommand{\ripe}{{RIPE\index{RIPE}}}

\newcommand{\glu}{{GLU\index{GLU}}}

\newcommand{\gipl}{{GIPL\index{GIPL}}}

\newcommand{\lucid}{{Lucid\index{Lucid}}}
\newcommand{\ilucid}{{Indexical Lucid\index{Indexical Lucid}}}
\newcommand{\jlucid}{{JLucid\index{JLucid}}}
\newcommand{\olucid}{{Objective Lucid\index{Tensor Lucid}}}
\newcommand{\tlucid}{{Tensor Lucid\index{Tensor Lucid}}}

\newcommand{\flucid}{{Forensic Lucid\index{Forensic Lucid}}}

\newcommand{\lucx}{{Lucx\index{Lucx}}}

\newcommand{\java}{{Java\index{Java}}}

\newcommand{\tab}[1]{\hspace{#1pt}}

\newcommand{\api}[1]{\texttt{#1}\index{API!#1}}

\newcommand{\lucidL}[1]{{$\mathit{Lucid}$}($L$) }

		{}

\def\myvert{\raise 2.27pt \hbox{\vrule depth 0pt height 8pt width 0.2mm}}
\def\myarrow{\hspace*{0.43mm}%
             \raise 2.29pt\hbox{\vrule depth 0pt height 8pt width 0.16mm}%
             \hspace*{-0.32mm}%
             $\longrightarrow$
             \ %
             }

\newcommand{\johndef}{\mathcal{D}}

\newcommand{\myid}{\textit{id}}

\newcommand{\mydagger}{\!\dagger\!}
\newcommand{\context}[2]{\mathcal{D},\mathcal{P} \vdash #1 : #2}

\newcommand{\qcontext}[2]{\mathcal{D},\mathcal{P} \vdash #1 \::\: #2}

\newcommand{\myifthenelse}{\mathtt{if}\;E\;\mathtt{then}\;E'\;\mathtt{else}\;E''}